\documentstyle[12pt,epsfig]{article}

\title{\  A Concordant ``Freely Coasting'' Cosmology}
\author{{Savita Gehlaut, Pranav Kumar}\\
       {Geetanjali and Daksh Lohiya\thanks{E--mail : dlohiya@iucaa.ernet.in}}\\
       {\em Department of Physics and Astrophysics,} \\
       {\em University of Delhi, Delhi-110007, India}\\
       {\em \&}\\
       {\em Inter University Centre for Astronomy and Astrophysics}\\
       {\em Ganeskhind, Pune 411 007, India}
       }

\begin {document}
\baselineskip=2\baselineskip
\maketitle 

\baselineskip= 15pt

\vskip 1cm
\centerline{\bf Abstract}

        A strictly linear evolution of the cosmological scale factor
is surprisingly an excellent fit to a host of cosmological observations.
Any model that can support such a coasting presents itself as a 
falsifiable model as far as classical cosmological tests are concerned.
Such evolution is known to be comfortably concordant with the Hubble 
diagram as deduced from data of recent supernovae 1a and high redshift objects,
it passes constraints arising from the age and gravitational lensing 
statistics and clears basic constraints on nucleosynthesis. 
Such an evolution exhibits distinguishable and verifiable features 
for the recombination era. This article discusses the concordance of 
such an evolution in relation to minimal requirements for 
large scale structure formation and cosmic microwave background anisotropy 
along with the overall viability of such models.

        While these results should be of interest for a host of alternative 
gravity models that support a linear coasting, we conjecture that a 
linear evolution would emerge naturally either from General Relativity itself
or from a General Relativistic theory of a non-minimally coupled scalar 
field theory. 

\vfil\eject

\section
{\bf Introduction}
\vspace{.5cm}

        Large scale homogeneity and isotropy observed in the universe
implies the Friedman-Robertson-Walker (FRW) metric:
\begin{equation}
\label{1}
ds^2 = dt^2 - a(t)^2[{dr^2\over {1 - Kr^2}} + r^2(d\theta^2 
+ sin^2\theta d\phi^2)]
\end{equation}
Here $K = \pm 1, 0$ is the curvature constant.
In standard ``big-bang'' cosmology, the scale factor $a(t)$ is
determined by the equation of state of matter and Einstein's equations. 
The scale factor, in turn, determines the response of a chosen 
model to cosmological observations. Four decades ago, the main 
``classical'' cosmological tests were (1) The
galaxy number count as a function of redshift; (2) The angular diameter
of ``standard'' objects (galaxies) as a function of redshift; and finally
(3) The apparent luminosity of a ``standard candle'' as a function of
redshift. Over the last two decades, other  tests 
that standard cosmology has been subjected to are: the early 
universe nucleosynthesis constraints, 
estimates of age of the universe in comparison to ages of old objects,
statistics of gravitational lensing, constraints from
large scale structure formation and finally, the physics of recombination
as deduced from cosmic microwave background anisotropy.

        In this article we explore concordance of the above observations 
with a FRW cosmology in which the scale factor evolves linearly with time:
$a(t) \propto t$, right from the creation event itself. 
The motivation for such an endeavor has been discussed at length in a series 
of earlier articles \cite{ddll}.
First of all, the use of Einstein's equations to describe cosmology has
never ever been justified. The averaging problem and the continuum limit
in General Relativity have not been properly addressed. In any case, most
treatments have not considered the retarded effects in their full 
generality\cite{ellis,ehlers,tavak}. Averaging Einsteins equations across
horizon lengths lacks self consistency. On the other hand,  
Newtonian cosmology, applied to an exploding {\it Milne ball}
in a flat space-time [see eg. 
\cite{milne,rindler}] gives a unique linear coasting cosmology viz. the FRW 
[{\it{Milne}}] metric with $a(t) = t$.
Such a cosmology does not suffer from the horizon problem. 
As a matter of fact, a linearly evolving model is the only power law model
that has neither a particle horizon nor a cosmological event horizon.
Linear evolution is also purged of the flatness or the fine tuning problem
\cite{ddll,dol,ford}. The scale factor in such theories
does not constrain the matter density parameter. The Linear coasting
characteristic of a Newtonian cosmology can be dynamically generated. As a 
matter of fact, it 
is a generic feature in a class of models that attempt to dynamically 
solve the cosmological constant problem \cite{wein,dol,ford,allen,mann}.
Particularly appealing in a fourth order conformally invariant model of 
Mannheim and Kazanas \cite{mann}. Gravitation is determined by a 
fourth order (Weyl tensor squared) action that identically vanishes for the
conformally flat FRW metric. A non-minimally coupled scalar field then produces
an effective \emph{repulsive gravitation} that quickly constrains the universe
to a linear coasting. Non-conformally flat perturbations around this linear
coasting background can be effectively described by eqn(3) below.

We may take any of the above as the basis for our linear coasting conjecture. 
In what follows, we assume that an homogeneous background FRW universe
is born and evolves as a Milne Universe about which a matter distribution
and standard General Relativity would determine the growth of perturbations. 
The following sections describe standard aspects of cosmology that follows.

\vspace{.5cm}
\section
{\bf A linearly coasting cosmology}
\vspace{.5cm}
\subsection
{\bf Classical Cosmology tests}

       Kolb\cite{kolb1} was probably first to demonstrate 
that data on Galaxy number counts as a function of red-shift as well as 
data on angular diameter distance as a function of red-shift do not rule
out a linearly coasting cosmology. Unfortunately, these two tests are 
marred by effects such as galaxy mergers and galactic evolution.  For
these reasons these tests have fallen into disfavour as reliable indicators of 
a viable model.

        The variation of apparent luminosity of a ``standard candle'' 
as a function of red-shift is referred to as the Hubble test.
With the discovery of Supernovae type Ia [SNe Ia] 
as reliable standard candles, the status of Hubble test has been elevated 
to that of a precision measurement.
Recent measurements by the supernovae cosmology project 
\cite{perl} eliminated the 
``minimal  inflationary'' prediction defined by $\Omega_{\Lambda} = 0$ and 
$\Omega_{M} = 1$. The data has been used to assess a ``non-minimal 
inflationary cosmology'' defined by $\Omega_\Lambda \ne 0$,~
$\Omega_{\Lambda} +\Omega_{M} = 1$.
The maximum likelihood analysis following from such a study has yielded 
the values $\Omega_M = 0.28 \pm 0.1$ and $\Omega_\Lambda = 0.72 \pm 0.1$
\cite{perl,wendy,branch}.

For a general power law cosmology with the scale factor 
$a(t) = {\bar k} t^\alpha$, with ${\bar k},~\alpha$
arbitrary constants, it is straightforward to discover 
the following relation between the apparent magnitude $m(z)$, the absolute
magnitude $M$  and the red-shift $z$ of an object:
\begin{equation}
m(z) = {\mathcal{M}} + 5logH_{o} + 5 log({\alpha \over H_o})^{\alpha}(1 + z) 
{\bar k}\emph{S}[{1 \over {(1 -\alpha){\bar k}}}
({\alpha \over H_o})^{1 - \alpha}( 1 - (1 + z)^{ 1- {1 \over \alpha}})] 
\end{equation}
Here $\emph{S}[X] = X, {\rm sin}(X)$ or ${\rm sinh}(X)$ for
$K = 0, \pm 1$ respectively, and ${\mathcal{M}} = M - 5log(H_o) + 25$. 
The best fit  turns out to be $\alpha = 1.001 \pm.0043$, $K = -1$.
\cite{ddll}. The minimum $\chi^2$ per degree of freedom turns out to be  1.18. 
This is comparable to the corresponding value
1.17 reported by Perlmutter et al for the constrained non-minimal inflationary 
cosmology. Linear coasting is as accommodating for 
more recent high red-shift objects as the standard non--minimal
inflationary model. The concordance of linear coasting with SNe1a data
finds a passing mention in the analysis of Perlmutter \cite{perl}
who noted that the curve for $\Omega_\Lambda = \Omega_M = 0$ (for which
the scale factor would have a linear evolution) is 
``practically identical to 
$\bf{best fit}$ plot for an unconstrained cosmology''. We display this
``practical coincidence'' in figure I that includes the more recent high 
redshift objects.

      The age estimate of the ($a(t) \propto t$) 
universe, deduced from a measurement of the Hubble parameter, is given
by $t_o = (H_o)^{-1}$.  The low red-shift SNe1a data \cite{ham,ham1}
gives the best value of $65~{\rm km~ sec^{-1}~ Mpc^{-1}}$ 
for the Hubble parameter.
The age of the universe turns out to be $15\times 10^9$ years.
This is $\approx$ 50\% greater than the age 
inferred from the same measurement in standard (cold) dark matter dominated
cosmology (without the cosmological constant). Such an age estimate is 
comfortably concordant with age estimates of old clusters. 

        A study of consistency of linear coasting with gravitational 
lensing statistics has recently been reported \cite{ddll}. The expected 
frequency of multiple image lensing events probes the 
viability of a given cosmology. A sample of 867 high luminosity 
optical quasars projected in a power law FRW cosmology gives an expected 
number of five lensed quasars for a linear coasting cosmology. This indeed
matches observations. Thus a strictly linear evolution of the scale factor
is comfortably concordant with gravitational lensing statistics.

\vspace{.2cm}
\subsection
{\bf ``The precision'' tests}
\vspace{.2cm}

        {\bf a) The Nucleosynthesis Constraint}
         What makes linear coasting particularly appealing is a recent
demonstration of  primordial nucleosynthesis 
not to be an impediment for a linear coasting cosmology 
\cite{ddll,steig}. 
A linear evolution of the scale factor may be expected to radically effect
nucleosynthesis in the early universe. Surprisingly, 
the following scenario goes through.
 
        Energy conservation, in a 
period  where the baryon entropy ratio does not change, enables 
the distribution of photons to be described 
by an effective temperature $T$ that scales
as $a(t)T = $ constant. 
With the age of the universe estimated from the Hubble parameter being
$\approx 1.5\times 10^{10}$
years, and $T_0 \approx 2.7K$, one concludes that the age of the
universe at $T \approx 10^{10}K$ would be some four
years [rather than a few seconds as in standard cosmology]. 
The universe would take some $10^3$ years to cool to $10^7K$. 
With such time periods being large 
in comparison to the free neutron life time, one would hardly expect any 
neutrons to survive at temperatures relevant for nucleosynthesis. 
However, with such a low rate of 
expansion, weak interactions remain in equilibrium for
temperatures as low as $\approx 10^8K$.
The neutron - proton ratio keeps falling as 
$n/p \approx exp[-15/T_9]$. Here $T_9$ is the temperature
in units of $10^9$K and the factor of 15 comes from the n-p mass 
difference in these units. There would again
hardly be any neutrons left if nucleosynthesis were to commence at (say)
$T_9 \approx 1$. However, as weak interactions are still in equilibrium, once
nucleosynthesis commences, inverse beta decay would replenish neutrons by
converting protons into neutrons and pumping them into the nucleosynthesis 
channel. With beta decay in equilibrium, the baryon entropy ratio determines 
a low enough nucleosynthesis rate that can remove neutrons out of the 
equilibrium buffer at a rate smaller than the relaxation time of the 
buffer. This ensures that neutron value remains unchanged as heavier nuclei
build up. It turns out that for baryon entropy ratio 
$\eta\approx 5\times 10^{-9}$, there would just
be enough neutrons produced, after nucleosynthesis 
commences, to give $\approx 23.9\% $ Helium and 
metallicity some $10^8$ times the metallicity produced in the early
universe in the standard scenario. This metallicity is 
of the same order of magnitude as seen in lowest metallicity objects.

        The only problem that one has to contend with is 
the significantly low yields of deuterium in such a cosmology. Though 
deuterium 
can be produced by spallation processes later in the history of the universe,
it is difficult to produce the right amount without a simultaneous over 
production of Lithium \cite{eps}.  However,
as pointed out in \cite{ddll}, the amount of Helium produced is quite 
sensitive to $\eta$ in such models. In an inhomogeneous universe, therefore,
one can have the helium to hydrogen ratio to have a large primordial 
variation. 
Deuterium can be produced by a spallation process much later in the history
of the universe. If one considers spallation of a helium 
deficient cloud onto a helium rich cloud, it is easy to produce deuterium
as demonstrated by Epstein \cite{eps} - without overproduction of Lithium.

        Interestingly, the baryon entropy ratio required for the right 
amount of helium corresponds to $\Omega_b \approx 0.2 $. Here $\Omega_b$ is
the ratio of the baryon density to a ``density parameter'' determined by the
Hubble constant: $\Omega_b \equiv \rho_b/\rho_c = 8\pi G \rho_b/3H_o^2$.
$\Omega_b \approx 0.2$ 
closes dynamic mass estimates of large galaxies and clusters
[see eg \cite{tully,peebls}]. 
In standard cosmology this closure is sought to be 
achieved by taking recourse to non-baryonic cold dark matter. Thus 
in a linearly scaling cosmology, there would be no need of non-baryonic cold 
dark matter to account for large scale galactic flows.

\vspace{.2cm}
{\bf b) Density Perturbations }
\vspace{.2cm}

We conjecture that the universe starts with a homogeneous isotropic 
distribution (eg. a Milne ball) and that an appropriate
averaging procedure that correctly takes care of retarded effects in a
relativistic gravity theory that accounts for a creation event, 
would support the large scale linear coasting with the metric 
described by eqn(1) with $K = -1$. 
Perturbations around this background shall be assumed to described by
\begin{equation}
-8\pi G \delta T^M_{\mu\nu} = \delta G_{\mu\nu}
\end{equation}
with $\delta G_{\mu\nu}$ determined by perturbations around eqn(1).
It would be fair to call this conjecture as a ``Newtonian Cosmology 
conjecture'' as Newton felt that one could justify the cancellation of
gravity in a homogeneous and isotropic universe. Bertschinger \cite{bert} and
Gibbons \cite{gwg} have shown how consisent cosmological dynamics, 
indistinguishable from dynamics in the standard model, follows from a 
Newtonian proposal in comoving rather than Minkowskian coordinates.
In effect this amounts to stating that {\it {a Homogeneous and isotropic 
universe with an arbitrary equation of state can be described by eqn(1), with
$a(t) = t, ~K = -1$ and that perturbations around it are determined by
eqn(3)}}. This further amounts to rejecting the unjustified assumptions 
of continuum limit (fluid approximation) and averaging that are usually 
taken for granted. Standard Einstein equations may well hold at the 
microscopic level but not as global averaged equations in a universe with a 
definite creation event. Further, as stated in the introduction, 
there are alternative gravity models that justify the above conjecture. 
In this article we have restricted ourselves to the consequenses of such a 
conjecture.

To explore whether such small perturbations in the above model can 
evolve to give structures at large scales, we express the metric as
\begin{equation}
ds^2 = ~^{(o)}g_{\mu\nu}dx^\mu dx^\nu + \delta g_{\mu\nu}dx^\mu dx^\nu
\end{equation}
with the background metric $^{(o)} g_{\mu\nu}$ expressed in comoving 
coordinates as:
\begin{equation}
^{(o)}g_{\mu\nu}dx^\mu dx^\nu = 
dt^2 - a(t)^2\gamma_{ij}dx^idx^j = a^2(\eta)(d\eta^2 -\gamma_{ij}dx^idx^j)
\end{equation}
$\eta$ is the conformal time ($d\eta = a^{-1}dt$, $a = t = t_0e^\eta$ ), and 
\begin{equation}
\gamma_{ij} = \delta_{ij}[1 - {1\over 4}(x^2 + y^2 +z^2)]^{-2}
\end{equation}
The $\delta g_{\mu\nu}$ describe the perturbation. These can be decomposed 
into scalar, vector or tensor type, depending on the way they transform on a
constant $\eta$ hypersurface. In the following, we consider scalar 
perturbations only. Following standard arguments,
vector perturbations can be shown to decay kinematically in an expanding 
universe and tensor perturbations lead to gravitational waves that do not 
couple to energy density and pressure inhomogeneities \cite{bardeen}.
%
%
Any scalar perturbation can be decomposed in terms of eigenmodes of
the laplacian on the constant $\eta$ surface:
\begin{equation}
\nabla^2Q \equiv \gamma^{ij}Q_{|ij} = -k^2Y
\end{equation}
where '$|$' represents a covariant derivative with respect to the
three metric $\gamma_{ij}$.
The most general form for the line element with scalar metric perturbations 
in the longitudinal gauge is (see eg. \cite{mukh,ruth})
\begin{equation}
ds^2 = a^2(\eta)[(1 + 2\Phi)d\eta^2 - (1 - \Psi)\gamma_{ij}dx^idx^j)]
\end{equation}
Similarly, perturbations of the stress tensor can be described in terms of
velocity and density perturbations in the longitudinal gauge 
($v^{(long)},~\delta^{(long)}$). We define the matter gauge
invariant variables as usual \cite{ruth} (the dot ``$^.$'' is a derivative
with respect to $\eta$):
\begin{equation}
V \equiv v^{(long)}
\end{equation}
\begin{equation}
D_g \equiv \delta^{(long)} + 3(1 + w)\Phi, ~~~~~~~ w \equiv p/\rho
\end{equation}
\begin{equation}
D \equiv \delta^{(long)} + 3(1 + w){\dot a\over a}{V\over k}
\end{equation}
For a perfect fluid, $\Phi = -\Psi$ and the full compliment of scalar
perturbation equations read:
\begin{equation}
4\pi Ga^2\rho D = (k^2 + 3)\Phi
\end{equation}
\begin{equation}
4\pi Ga^2(\rho + p)V = k(({{\dot a}\over a})\Psi - \dot\Phi)
\end{equation}
\begin{equation}
\dot D_g + 3(c_s^2 - w)D_g\dot a/a + (1 + w)kV + 3w\Gamma\dot a/a = 0
\end{equation}
\begin{equation}
\dot V + (1 - 3c_s^2)V{\dot a\over a} = k(\Psi - 3c_s^2\Phi) + 
{{c_s^2k}\over {1 + w}}D_g + {{wk}\over {1 + w}}\Gamma   
\end{equation}
where $c_s$ is the sound speed and $\Gamma \equiv \Pi_l - c_s^2\delta/w$
vanishes for adiabatic perturbations (see eg. \cite{ruth,mukh}).
For pure dust $w = c_s^2 = p = 0$ and the above equations reduce to:
\begin{equation}
\dot D_g + kV = 0 
\end{equation}
\begin{equation}
\dot V + V{\dot a\over a} = k\Psi  
\end{equation}
\begin{equation}
-k^2\Psi = 4\pi Ga^2\rho (D_g + 3(\Psi + ({\dot a\over a}){V\over k}))  
\end{equation}
In terms of the present density,
$$
C \equiv 4\pi G \rho_o a_0^2 = {3\over 2}{{8\pi G}\over {3H_o^2}}\rho_o 
= {3\over 2}\Omega_b \approx 0.3~~ {\rm {for}}~~ \Omega_b \approx 0.2 
$$
the density perturbation equation simply reduces to:
\begin{equation}
[(k^2 + 3){{e^\eta}\over C} + 3]\ddot D_g +
[(k^2 + 3){{e^\eta}\over C} + 2]\dot D_g
- D_gk^2  = 0 
\end{equation}
$k = 1$ corresponds to the Hubble scale which is the same as the curvature
scale in this model. At a redshift $\approx 1000$, this scale subtends an
angle roughly .25 degrees. Using constraints from microwave background 
anisotropy at these angles gives $D_g \approx 10^{-5}$ at these scales at
the last scattering surface. It is easy to see that modes 
$k \stackrel{<}{\sim} 1$ do not grow. At smaller angular scales (large $k$),  
the observed anisotropy is expected to fall to much lower values 
\cite{savthes}. 
Diffusion damping dampens anisotropies at angular scales smaller than  
about one minute. However, for such large values of $k$, $D_g$ 
has rapidly growing solutions. Eqn.(19) goes as 
\begin{equation}
\ddot D_g + \dot D_g - Ce^{-\eta}D_g = 0  
\end{equation}
This has exact solutions in terms of modified first and second type 
bessel functions $I_1,~K_1$:
\begin{equation}
D_g = C_1(Ce^{-\eta})^{1\over 2}I_1((4Ce^{-\eta})^{1\over 2})
       + C_2(Ce^{-\eta})^{1\over 2}K_1((4Ce^{-\eta})^{1\over 2})
\end{equation}
For large arguments, these functions have their asymptotic forms:
\begin{equation}
I_1 \longrightarrow {(Ce^{-\eta})^{-{1\over 4}}\over {2\sqrt{\pi}}}
exp[2(Ce^{-\eta})^{1\over 2}];~~
K_1 \longrightarrow {(Ce^{-\eta})^{-{1\over 4}}\over {2\sqrt{\pi}}}
exp[- 2(Ce^{-\eta})^{1\over 2}]
\end{equation}
Even if diffusion damping were to reduce the baryon density contrast to 
values as low as some $10^{-15}$, a straight forward numerical integration
of eqn(19) demonstrates that for $k \ge 3000$ the density contrast becomes 
non linear around redshift of the order 50. 

In contrast to the above, in the radiation dominated epoch 
$ w = c_s^2 = 1/3$. In adiabatic approximation, eqns(12,15) imply:
\begin{equation}
[(k^2 + 3){3\over {4k^2}} + {3\tilde C\over {2k^2e^{2\eta}}}]\ddot D_g +
{3\tilde C\over {k^2e^{2\eta}}}\dot D_g
+[{{k^2 + 3}\over 8} - {\tilde C\over {2e^{2\eta}}}]D_g  = 0 
\end{equation}
For $\eta$ large and negative, small $k$ pertubation equation reduces to 
\begin{equation}
3\ddot D_g + 6\dot D_g -k^2D_g = 0
\end{equation}
It is straightforward to demonstrate that 
perturbations bounded for large negative $\eta$,  damp out for small $k$.
It is more instructive to work with the equation for the potential for 
large $k$ perturbations:
\begin{equation}
\ddot \Phi + 3(1 + c_s^2)\dot\Phi + [2(1 + 3c_s^2) + k^2c_s^2]\Phi = 0
\end{equation}
when 
\begin{equation}
\delta^{long} = {{2e^{2\eta}}\over {3\Omega_\gamma}}[ 3\dot\Phi + k^2\Phi]
\end{equation}
For $c_s = 1/3$, the solutions are 
\begin{equation}
\Phi \propto e^{-2\eta}e^{\pm({{ik\eta}\over {\sqrt{3}}})}
\end{equation}
\begin{equation}
\delta^{long} \propto (k^2 \pm \sqrt{3}ik)e^{\pm({{ik\eta}\over {\sqrt{3}}})} 
\end{equation}
Thus we get oscillating solutions for the potential as well as the 
density fluctuations. These fluctuations shall manifest themselves as 
``acoustic'' fluctuations in the Cosmic microwave background sky to which 
we shall return later.

We summarize here by noting that fluctuations do not grow in the radiation 
dominated era, small $k$ (large scale) fluctuations do not grow in the
matter dominated era as well, however, 
even tiny residual fluctuations $O(10^{-15})$ 
at the last scattering surface for large values of $k \ge 3000$ 
in the matter dominated era, can in principle grow to the non linear regime. 
Such a growth would be a necessary condition for structure formation and is
not satisfied in the standard model. In the standard model, cold dark matter
is absolutely essential for growth. Thus a linear coasting model does not
need such dark matter to have small density perturbations approach a 
non linear regime necessary for a structure formation theory.

\vspace{.2cm}
{\bf c) The recombination epoch}
\vspace{.2cm}

        Salient features of the plasma era in a linear coasting cosmology 
have been described in \cite{savitaI,savthes}. Here we reproduce some of
the peculiarities at the recombination epoch. These
are deduced by making a simplifying assumption of thermodynamic 
equilibrium just before recombination. 

A recombination process 
that directly produces a Hydrogen atom in the ground state releases a 
photon with energy $B = 13.6 eV$ in each recombination. $n_\gamma(B)$,
the number density of photons in the background radiation with energy $B$, 
is given by [see eg. \cite{seager}]:
\begin{equation}
{{n_\gamma (B)}\over n} = {{16\pi}\over n}T^3exp({{-B}\over T})
\approx {{2\times 10^6}\over {h^2}}exp({-{13.6}\over \tau})
\end{equation}
Where $\tau$ is the temperature in units of eV.
This ratio is unity at $\tau \approx .9 $ for $\Omega_Bh^2 \approx 1$
 and decreases rapidly at lower 
temperatures. Any 13.6 eV photons released due to recombination 
have a high probability of ionizing neutral atoms formed a little earlier. 
This process is therefore not very effective for producing a net number of 
neutral atoms. The way out of this empasse is the {\emph{two 
photon emmission}}.
The $2S\longrightarrow 1S$ transition is strictly forbidden at first order. 
The conservation of both the energy and angular momentum in the transition 
can take place only by emmission of a pair of photons. This gives a mechanism
for transferring the ionization energy into photons with $\lambda > 
\lambda_{Ly\alpha}$. Since the reverse process does not occur at the same rate,
this is non-equilibrium recombination. Being a process of second order in
perturbation theory, this is slow (lifetime $\approx$ 0.1 sec). 
As recombination has to pass through this bottleneck, it proceeds 
with a rate different from the Saha preediction.

In the red-shift range $800 < z < 1200$, the approximate fractional
ionization in a linearly coasting cosmology can be approximated by
\cite{savthes}:
\begin{equation}
x_e = {{7.6\times 10^{-5}}\over {\Omega_bh}}({z\over {1000}})^{12.25}
\end{equation}
The optical depth for Thompson scattering is then:
\begin{equation}
\tau_\gamma = \int_0^t n_B(t)x_e(t)\sigma_T dt
= -\int_0^zn_B(z)x_e(z)\sigma_T({{dt}\over {dz}})dz
\end{equation}
With $n_b(z) = \eta n_\gamma(z) = \eta\times 421.8(1+z)^3 ~cm^{-3}$, and
\begin{equation}
{{dt}\over {dz}} = - {1\over {H_0(1+z)^2}}
\end{equation}
one can find the red-shift at which the optical depth goes to unity. 
With the residual ionization $x_{e,res}$, we get
\begin{equation}
\tau_\gamma = 0.55 ({z\over {1000}})^{14.25}
\end{equation}
From this optical depth, we can compute the probability that a photon was
last scattered in the interval $(z,z + dz)$. This is given by:
\begin{equation}
P(z) = e^{-\tau_\gamma}{{d\tau_\gamma}\over {dz}} \approx 7.85\times 10^{-3}
({z\over {1000}})^{13.25}exp[-0.55({z\over {1000}})^{14.25}]
\end{equation}
This $P(z)$ is sharply peaked and well fitted by a gaussian of mean redshift
$z \approx 1037$ and standard deviation in redshift $\Delta z \approx 67.88$.
Thus in a linearly coasting cosmology, the last scattering surface locates at
redshift $z^* = 1037$ with thickness 
$\Delta z \approx 68$. Corresponding values in 
standard cosmology are $ z = 1065$ and $\Delta z \approx 80$ \cite{paddy}.

   An important scale that determines the nature of CMBR anisotropy is the 
Hubble scale which is the same as the curvature scale for linear coasting. 
The angle subtended today, by the Hubble radius at $z^* = 1037$, is 
determined by the angular diameter - redshift relation:
\begin{equation}
r_R {\rm sin}{\theta\over 2} = {\rm sinh}[{{d(\theta)(1 + z_R)}\over {2a_o}}]
\end{equation}
For $d(\theta) = 2d_H(t_R) = 2H(t_R)^{-1} = 2[H_o(1 + z_R)]^{-1}$ and
$r_R \approx (1 + z)/2$, this gives:
\begin{equation}
({{1+z_R}\over 2}){\theta\over 2} ={\rm sinh(1)}
\end{equation}
or $\theta_H \approx 15.5$ minutes. 

        In standard cosmology, the {\it sound horizon} is of the same order as 
the Hubble length at recombination. The Hubble length determines the scale 
over which physical processes can occur coherently. Thus, as we suggest in the
next section, one expects all 
acoustic signals to be contained within an angle of the order of the angle 
subtended by the Hubble length at recombination.

In a linear coasting, the Hubble length is 
precisely the inverse of the curvature scale. However, the sound horizon 
($s^*$) is
much larger. Strictly speaking, the particle as well as the sound horizon are
infinite for a linear coasting cosmology. As we would see, the CMB anisotropy
pattern frozen on the Last scattering surface depends on the distance 
travelled by pressure waves from the epoch baryon density starts becoming 
comparable to radiation density.
The photon diffusion scale is determined by the thickness of the LSS.
With $z^* \approx 1037$ and $\Delta z \approx 68$, the photon diffusion scale
projected on the LSS corresponds to an angular size roughly one thirteenth of 
the Hubble length at the LSS. This subtends an angle of roughly one minute 
at the current epoch.

        The above scales in principle determine the nature of CMB anisotropy.
The CMB effectively ceases to scatter when the optical
depth to the present drops to unity. After last scattering, the photons 
effectively free stream. On the LSS, the photon
distribution may be locally isotropic while still possessing inhomogeneities
i.e. hot and cold spots, which will be observed as anisotropies in the 
sky today (see eg. \cite{hu,tegmark}). 

\vspace{.2cm}
{\bf d) The CMB temperature anisotropy}
\vspace{.2cm}

The fractional temperature fluctuation $\Theta(t,{\bf x}, \gamma)$ 
for a black body is defined as
\begin{equation}
4\Theta \equiv {\delta \rho_{\gamma} \over \rho_{\gamma}}
\end{equation}
where $\rho_{\gamma} \propto T^4$ is the spatially and directionally 
averaged energy density of the photons. $\Theta$ satisfies the 
collisionless Boltzmann equation \cite{hu,savthes}:
\begin{equation}
{d\over d\eta}[\Theta + \Phi](\eta,{\bf x}, \gamma) = \dot\Phi + \dot\Psi
\end{equation}

        Before recombination photons and baryons were strongly coupled
via Compton scattering. In this case the collisional Boltzmann equation
is given by
\begin{equation}
{d \over d\eta}(\Theta + \Phi) \equiv
\dot\Theta + \gamma^i {\partial \over \partial x^i}(\Theta + \Phi) +
\dot\gamma^i {\partial \over \partial \gamma^i}\Theta - \dot\Psi = 
\dot\tau (\Theta_0 - \Theta - \gamma_i v^i_b + {1\over 16}
\gamma_i \gamma_j \Pi^{ij}_{\gamma})
\end{equation}
where $\dot\tau \equiv x_e n_e \sigma_T$ is the differential optical depth
to Thomson scattering with $\sigma_T = 8\pi \alpha^2/3m^2_e$ as the
Thomson scattering Cross-section, $x_e$ and $n_e$ are the ionization 
fraction and total electron density respectively. 
$\Theta_0 = \delta_\gamma /4$
is the isotropic component of $\Theta$, $\gamma^i = \dot x^i$,
and the quantities $\Pi^{ij}_{\gamma}$ are the quadrupole moments of the 
energy distribution given by
\begin{equation}
\Pi^{ij}_{\gamma} = \int {d\Omega \over 4\pi}
(3\gamma^i \gamma^j - \gamma^{ij}) 4\Theta(\eta,{\bf x}, \gamma)
\end{equation}

Perturbations are expanded in terms of eigenfunctions
of the Laplacian. To effect this,
spherical coordinates are defined by the the line element
\begin{equation}
dl^2 = \gamma_{ij}dx^i dx^j = -K^{-1} [d\chi^2 + {\rm sinh}^2\chi (d\theta^2
+ {\rm sin}^2\theta d\phi^2)]
\end{equation}
where the distance is scaled to the curvature radius $\chi = \sqrt{-K}\eta$.
The laplacian of any scalar function $Q$ is given by:
\begin{equation}
\gamma^{ij}Q_{|ij} = -K {\rm sinh}^{-2}\chi [{\partial\over \partial \chi}
({\rm sinh}^{2}\chi {\partial Q \over \partial\chi}) + {\rm sin}^{-1}\theta
{\partial\over \partial \theta}
({\rm sin}\theta {\partial Q \over \partial\theta}) + {\rm sin}^{-2}\theta
{\partial^2 Q\over \partial \phi^2} ]
\end{equation}
Since the angular part is independent of curvature, we may separate 
variables such that $Q =  X^l_{\nu}(\chi)Y^m_l(\theta,\phi)$ and
$\nu^2 \equiv -(k^2/K + 1)$. The spherically symmetric $l = 0$ function 
turns out to be:
\begin{equation}
X^0_\nu(\chi) = {{\rm sin}(\nu\chi) \over \nu {\rm sinh}\chi}
\end{equation}
The higher modes are explicitely given by \cite{Abbot}
\begin{equation}
 X^l_{\nu}(\chi) = (-1)^{l + 1} M^{-1/2}_l \nu^{-2} (\nu^2 + 1)^{-l/2}
{\rm sinh}^l\chi {d^{l + 1}({\rm cos}\nu\chi) \over d({\rm cosh}\chi)^{l + 1}}
\end{equation}
which reduce to $j_l(k\eta)$ in the flat space limit, where
$$M_l \equiv \Pi^l_{l' = 0}K_{l'}$$
$$K_0 = 1$$
\begin{equation}
K_l = 1 - (l^2 - 1)K/k^2,~~~~~~~~l\ge 1.
\end{equation}
All these reduce to unity as $K \rightarrow 0$. The functions are normalized 
as \cite{hu,lyth}
\begin{equation}
\int X^l_{\nu}(\chi) X^{l'}_{\nu'}(\chi){\rm sinh}^2\chi d\chi = 
{\pi \over 2\nu^2} \delta(\nu - \nu') \delta(l - l')
\end{equation} 
These functions can also be generated from their recursion relations.
One particularly useful relation, which is used in the collisionless
Boltzmann equation, is \cite{Abbot}
\begin{equation}
{d \over d\eta}X^l_{\nu} =  {l\over 2l + 1}kK^{1/2}_l X^{l - 1}_{\nu}
+ {l + 1 \over 2l + 1}kK^{1/2}_{l + 1} X^{l + 1}_{\nu}
\end{equation}

        Vectors and tensors needed in the
description of the velocity and stress perturbations can be constructed
from the the covariant derivatives of the scalar function $Q$ and the
metric tensor,
$$Q_i \equiv -k^{-1} Q_{|i}$$
\begin{equation}
Q_{ij} \equiv k^{-2} Q_{|ij} + {1\over 3} \gamma_{ij} Q
\end{equation}
where the indices are to be raised and lowered by the three metric 
$\gamma^{ij}$ and $\gamma_{ij}$. These vectors and tensors can be 
used to represent dipoles and quadrupoles, $G_1 = \gamma^i Q_i$
and $G_2 = {3\over 2}\gamma^i \gamma^j Q_{ij}$. In general any
function of position {\bf x} and angular direction  ${\mathbf\gamma}$,
e.g. the radiation distribution function, can be represented as \cite{Wilson}
\begin{equation}
F({\bf x}, {\bf \gamma}) = \sum_{\tilde {\bf k}}\sum^{\infty}_{l = 0}
\tilde F_l({\bf k}) G_l({\bf x}, {\bf \gamma}, {\bf k})
\end{equation}
where
\begin{equation}
G_l({\bf x}, {\bf \gamma}, {\bf k}) = (-k)^{-l} Q_{|i_1...i_l}({\bf x},{\bf k})
P^{i_1...i_l}_l ({\bf x}, {\bf \gamma})
\end{equation}
and $\tilde k^2 = K(k^2/K + 1)$ and
$$P_0 = 1,~~~~~~~~~~~~~P^i_1 = \gamma^i$$
$$P^{ij}_2 = {1\over 2}(3\gamma^i\gamma^j - \gamma^{ij})$$
\begin{equation}
P^{i_1...i_l}_{l + 1} = {2l + 1 \over l + 1}\gamma^{(i_1}P^{i_2...i_{l+1})}_l
 -  {l  \over l + 1}\gamma^{(i_1i_2}P^{i_3...i_{l+1})}_{l - 1}
 \end{equation}
with parentheses denoting symmetrization about the indices.

The multipole decomposition of $\Theta$ reads:
\begin{equation}
\Theta (\eta, {\bf x}, {\bf \gamma}) = \sum^{\infty}_{l = 0}
 \Theta_l({\eta}) M^{-1/2}_l G_l({\bf x}, {\bf \gamma})
\end{equation}
 Recursion relations for $G_l$ give the 
standard hierarchy of coupled equations for the $l-$modes:
$$\dot\Theta_0 = -{k \over 3}\Theta_1 + \dot\Psi,$$
$$\dot\Theta_1 = k[\Theta_0 + \Phi - {2\over5}K^{1/2}_2 \Theta_2] 
- \dot\tau(\Theta_1 - V_b),$$
$$\dot\Theta_2 = k[{2\over 3}K^{1/2}_2\Theta_1  - {3\over7}K^{1/2}_3 \Theta_3] 
- {9\over 10}\dot\tau\Theta_2, $$
\begin{equation}
\dot\Theta_l = k[{l\over 2l - 1}K^{1/2}_l\Theta_{l - 1}  - {l +1\over2l + 3}
K^{1/2}_{l + 1} \Theta_{l + 1}] - \dot\tau\Theta_l,~~~~(l>2)
\end{equation}
where $ \gamma_i v^i_b({\bf x}) = V_b G_1({\bf x}, {\bf\gamma})$ and 
$K_l = 1 - (l^2 - 1)K/k^2$. 

Before recombination,
Compton scattering transfers momentum between the photons and baryons.
From the conservation of momentum of the photon baryon fluid, we get 
$$\dot\delta_b = -kV_b + 3\dot\Psi$$
\begin{equation}
\dot{V}_b = -{\dot{a}\over a}V_b + k\Phi + \dot\tau(V_\gamma - V_b)/R
\end{equation}
where $R = 3\rho_b /4\rho_\gamma$. 
The baryon continuity equation 
can also be combined with the photon continuity equation
to obtain \cite{hu}
\begin{equation}
\dot\delta_b = -k(V_b - V_\gamma) + {3\over 4}\dot\delta_\gamma
\end{equation}

The differential optical depth $\dot\tau$ is high enough before recombination
making Compton scattering extremely rapid and effective. 
Since photon-baryon tight coupling approximation holds, the Boltzmann 
eqn(53) and the Euler eqn(54) for baryons can be expanded in the 
Compton scattering time $\dot\tau^{-1}$. To zeroth order, we get the 
tight-coupling identities,
$$\dot\Theta_0 = {1\over 3}\dot\delta_b$$  
$$\Theta_1 \equiv V_\gamma = V_b$$
\begin{equation}
\Theta_l = 0~~~~~~~~~l~\ge~2
\end{equation}
These equations merely express the fact that the radiation is isotropic in 
the baryon rest frame and the density fluctuations in the photons grow 
adiabatically with the baryons. Substituting the zeroth order solutions 
back into equations (53) and (54), we obtain the iterative first order 
equations:
$$\dot\Theta_0 = -{k\over 3}\Theta_1 + \dot\Psi$$
 \begin{equation}
\dot\Theta_1 = -{\dot{R}\over 1 + R}\Theta_1 + {1\over 1 + R}k\Theta_0 + k\Phi
\end{equation}
where we have used the relation $\dot{R} = (\dot{a}/a)R$. The tight coupling 
approximation eliminates the multiple time scales and the infinite hierarchy 
of coupled equations of the full problem. These equations can be combined to 
form a single second order equation,
\begin{equation}
\ddot\Theta_0 + {\dot{R}\over 1 + R}\dot\Theta_0 + k^2c^2_s \Theta_0 = F
\end{equation}
where the photon-baryon sound speed is 
\begin{equation}
c^2_s \equiv {\dot{p}_\gamma \over \dot\rho_\gamma + \dot\rho_b} = 
{1\over 3}{1\over 1 + R}
\end{equation}
assuming that $p_b \approx 0$ and 
\begin{equation}
F = \ddot\Psi + {\dot{R}\over 1 + R}\dot\Psi -{k^2 \over 3}\Phi
\end{equation}
is the forcing function.

Essential features of CMB anisotropy follow from the above analysis. 
In the rest of this article we shall be content with making 
rough estimates of CMB anisotropy peak locations to judge whether there 
is any a-priori discordance with observations.

Let us first consider ignoring the time dependence of
the potentials $\Psi$ and $\Phi$ and also the baryon-
photon momentum ratio $R$. Then eqn(58) reduces to:
\begin{equation}
\ddot\Theta_0  + k^2c^2_s \Theta_0 = -{k^2 \over 3}\Phi
\end{equation}
This is a simple harmonic oscillator under the constant acceleration provided 
by gravitational infall and its solution is:
\begin{equation}
\Theta_0(\eta) = [\Theta_0(0) + (1 + R) \Phi]\rm{cos}(kr_s) + {1\over kc_s}
\dot\Theta_0(0)\rm{sin}(kr_s) - (1 + R)\Phi
\end{equation}
where $r_s$ is the sound horizon given as  
\begin{equation}
r_s = \int^{\eta}_{\eta_0} c_sd\eta'
\end{equation}
where $\eta_0$ is the epoch of birth of pressure waves.
We may take this to be the QGP phase transition epoch 
$T_{QGP} \approx 10^{12}K$, or even earlier. 
The following results are independent of this initial epoch (cut-off).
The distance a sound wave travels from the initial epoch till the epoch 
$\eta$ is $r_s$.  $\Theta_0(0)$ and $\dot\Theta_0(0)$ govern the form of the 
acoustic oscillations and govern the adiabatic
and isocurvature modes respectively.
Eqn(62) also implies, through the photon continuity eqn(57), that
\begin{equation}
\Theta_1(\eta) = 3[\Theta_0(0) + (1 + R) \Phi]c_s \rm{sin}(kr_s) + 3{1\over k}
\dot\Theta_0(0)\rm{cos}(kr_s) 
\end{equation}
In eqns(62, 64) lie the main acoustic and redshift effects 
which dominate primary anisotropy formation.
In the early universe, photons dominate the fluid and $R \rightarrow 0$.
In this limit, the oscillation takes on an even simpler form. For the adiabatic
mode, $\dot\Theta_0(0) = 0$ and 
$\Theta_0(\eta) = [\Theta_0(0) + \Phi]\rm{cos}(kr_s) - \Phi$. This represents 
an oscillator with a zero point which has been 
displaced by gravity. The zero point represents the state at which gravity 
and pressure are balanced. The displacement $-\Phi > 0$ yields hotter photons 
in the potential well since gravitational infall not only increases the
 number density of the photons but also their energy through gravitational
 blueshifts.

        However, photons also suffer a gravitational redshift from climbing
 out of the potential well after last scattering. This precisely cancels the  
$-\Phi$ blueshift. Thus the effective temperature perturbation is 
$\Theta_0(\eta) + \Phi = [\Theta_0(0) + \Phi]\rm{cos}(kr_s)$.
This takes care of two sources of primary anisotropy: gravity and intrinsic
temperature variations on the last scattering surface. The final source in
this (tight coupling) approximation is the Doppler shift. 
The bulk velocity of the fluid along the line of sight is $V_\gamma/\sqrt{3}
= \Theta_1/\sqrt{3}$. This causes the observed temperature to be 
Doppler shifted. Eqn(64) shows that
the acoustic velocity $\Theta_1$ is $\pi/2$ out of phase with the temperature.
Because of its phase relation, the velocity contribution  fills in the 
zeros of the temperature oscillations. The rms temperature fluctuation is
identically smoothened out leaving no residual primary anisotropies.

This null result gets altered once we take into account of a 
non-vanishing $R$. Baryons add gravitational and inertial mass to the fluid, 
$m_{eff} = 1 + R$, without affecting the pressure. This decreases the sound
speed and changes the balance of pressure and gravity. Gravitational infall
now leads to greater  compression of the fluid in a potential well, i.e. a 
further displacement of the oscillation zero point. Since the redshift is 
not affected by the baryon content, this relative shift remains after last 
scattering to enhance all peaks from compression over those from rarefaction.
To demonstrate the effect of $R \ne 0$, let us assume it to be constant and
non-vanishing after an epoch $\eta = \eta_1$ and negligible earlier. 
For an adiabatic perturbation, $\dot\Theta_0(0) = 0$ we have:
\begin{equation}
\Theta(\eta_1) = [\Theta(0) + \Phi]\rm{cos}({k\over {\sqrt{3}}}\eta_1) - \Phi
\end{equation}
\begin{equation}
\dot\Theta(\eta_1) = - {k\over {\sqrt{3}}}[\Theta(0) + \Phi]
\rm{sin}({k\over {\sqrt{3}}}\eta_1) 
\end{equation}
Thus after $\eta = \eta_1$, the solution reads:
$$
\Theta(\eta) + (1 + R)\Phi 
= [\Theta(0) + \Phi][\rm{cos}({k\over {\sqrt{3}}}\eta_1)
\rm{cos}({k\over {\sqrt{3(1 + R)}}}(\eta - \eta_1))$$
\begin{equation}
- \sqrt(1 + R)\rm{sin}({k\over {\sqrt{3}}}\eta_1)
\rm{sin}({k\over {\sqrt{3(1 + R)}}}(\eta - \eta_1))]
+ R\Phi \rm{cos}({k\over {\sqrt{3(1 + R)}}}(\eta - \eta_1))
\end{equation}

        The phase of the oscillation freezes at the last scattering 
surface $\eta = \eta^*$ and determines the observed fluctuation pattern. 
For a linear coasting cosmology, $|\eta_1|$ is large. It is in fact infinite
for a universe starting at $t = 0$, and can be made finite (but large) only by 
choosing a lower cutoff for $t$. The angular power spectrum of the above
expression is obtained by transforming it with respect to the ultraspherical
bessel functions \cite{hu,lyth}. A given $k$ mode would contribute to two
distinct angular scales: (1) A very small angular scale corresponding to
the large phase of the first two harmonic functions, and (2) the angular 
scale corresponding to the third harmonic function:
$$
+ R\Phi \rm{cos}({k\over {\sqrt{3(1 + R)}}}(\eta^* - \eta_1))
$$
For $\Omega_b \approx 0.2$ and $h \approx 0.7$, significant deviation of 
$c_s$ from $1/\sqrt{3}$ occur around the matter radiation equality at 
redshift given by $1 + z_{eq} \approx 3.9\times 10^4(\Omega_bh^2)\approx 4000$.
Thus only those fluctuations that originate at redshift $\tilde z = \rm{few}
\times z_{eq}$ would give rise to observable temperature anisotropies at LSS
which locates at $z^* \approx 1037$.
The argument of the above harmonic pattern freezes at LSS and would give rise 
to ``peaks'' at
\begin{equation}
{k\over {\sqrt{3(1 + R)}}}a_{Rec}\rm{ln}({\tilde z\over z_r}) = n\pi
\end{equation}
The corresponding wavelengths are at
\begin{equation}
\lambda^a_n = {\pi\over k_n}
\end{equation}
The first ``peak'' is at $k_a = 0$ and the second at roughly $\theta \approx
15$ minutes. Compression peaks occur at odd values of $n$ and are located
at angles $\theta^a_n = 15/n\pi$ minutes. The entire analysis can be repeated
for isocurvature modes to find the peaks located at even $n$ at 
$\theta^i_n = 15/(n +{1 \over 2})$ minutes. 

       The above analysis has not taken into account the variation of $R$. 
For a qualitative comparison one has to integrate eqn(58). Using a WKB
approximation (which is good for modes for which the time scale of variation
of sound speed is greater than the oscillation period) it can be seen
\cite{hu} that the phase structure only suffers a minor change from the above
analysis while most of the effect of changing $R$ effects the amplitude of the
solution.

        The exact profile of the anisotropy would be determined by the choice
of the nature of initial conditions (adiabatic or isocurvature), the chosen 
initial power spectrum, and the growth of perturbations after $z^*$ 
(these determine the late or the {\it integrated SW effect}), 
aspects of reionization etc. 
As photons possess a mean free path in the Baryons 
$\lambda_c \approx \dot\tau^{-1}$ due to Compton scattering, the
 photon-baryon tight coupling breaks down at the photon diffusion scale.
Finally, after last scattering, photons free stream toward the observer on 
radial null geodesics and suffer only the gravitational interactions of 
redshift and dilation. Spatial fluctuations on the last scattering surface 
are observed as anisotropies in the sky. Free streaming thus transfers
 $l = 0$ inhomogeneities and $l =1$ bulk velocities to high multipoles as 
the $l-$mode coupling of the Boltzmann eqn(53) suggests. Isotropic 
$l = 0$ density perturbations are thus averaged away collisionlessly.

\vspace{.5cm}
\section
{\bf Summary}
\vspace{.5cm}

        The main point we make in this article is that in spite
of a significantly different evolution, the recombination history of a 
linearly coasting cosmology can be expected to give the location of 
the primary acoustic peaks in the 
same range of angles as that given in Standard Cosmology. 
Given that none of the alternative anisotropy formation scenarios 
provide a compelling {\it ab initio} model  \cite{hu1995} , 
it is perhaps best to keep an open 
mind to all possibilities. As the large scale structure and CMB anisotropy 
data continue to accumulate, one could explore the general principles 
for an open coasting cosmology to aid in the empirical 
reconstruction of a consistent model for structure formation.

        Finally, we are tempted to mention that a linear coasting cosmology
presents itself as a falsifiable model. It is encouraging to observe its
concordance !! In standard cosmology, falsifiability has taken on a backstage
- one just constrains the values of cosmological parameters subjecting 
the data to Bayesian statistics. Ideally, one would have been very content
with a cosmology based on physics that we have already 
tested in the laboratory.
Clearly, standard cosmology does not pass that test. One needs a mixture of hot
and cold dark matter, together with (now) some form of \emph{dark energy} 
to act as a cosmological constant, to find any concordance with observations. 
In other words, one uses observations to parametrize theory in Standard 
Cosmology. 

        Linear coasting presents a very distinguishable cosmology. 
Recombination occurs at age $\sim 10^7$ years as opposed to $\sim 10^5$
years in standard cosmology. The Hubble scale at decoupling is thus two
orders of magnitude greater in linear coasting. This fact, coupled with
the absence of any horizon, could well have falsified linear coasting. Any
concordance with observations is therefore very significant. The model
comes with its characteristic predictions. Thus linear coasting
has the potential of relegating the need for any form of dark matter or 
dark energy (or for that matter, any physics not already tested in the 
laboratory) to the physics archives where they enjoy the same status as
ether and phlogiston. The message this article is to convey is that a universe
that is born and evolves as a curvature dominated model has a tremendous
concordance and there are sufficient grounds to explore models that support
such a coasting.

\vfil
\eject 

\vskip 0.5 cm
{\bf Acknowledgments}

We thank the University Grants Commission for financial support.
The work benefited from discussions with Profs. T. Padmanabhan,  
K. Subrahmanian and T. Saurodeep and the same are gratefully acknowledged.
\vskip 1cm

\bibliography{plain}

\vfil
\eject

\begin{figure}
\epsfig{file=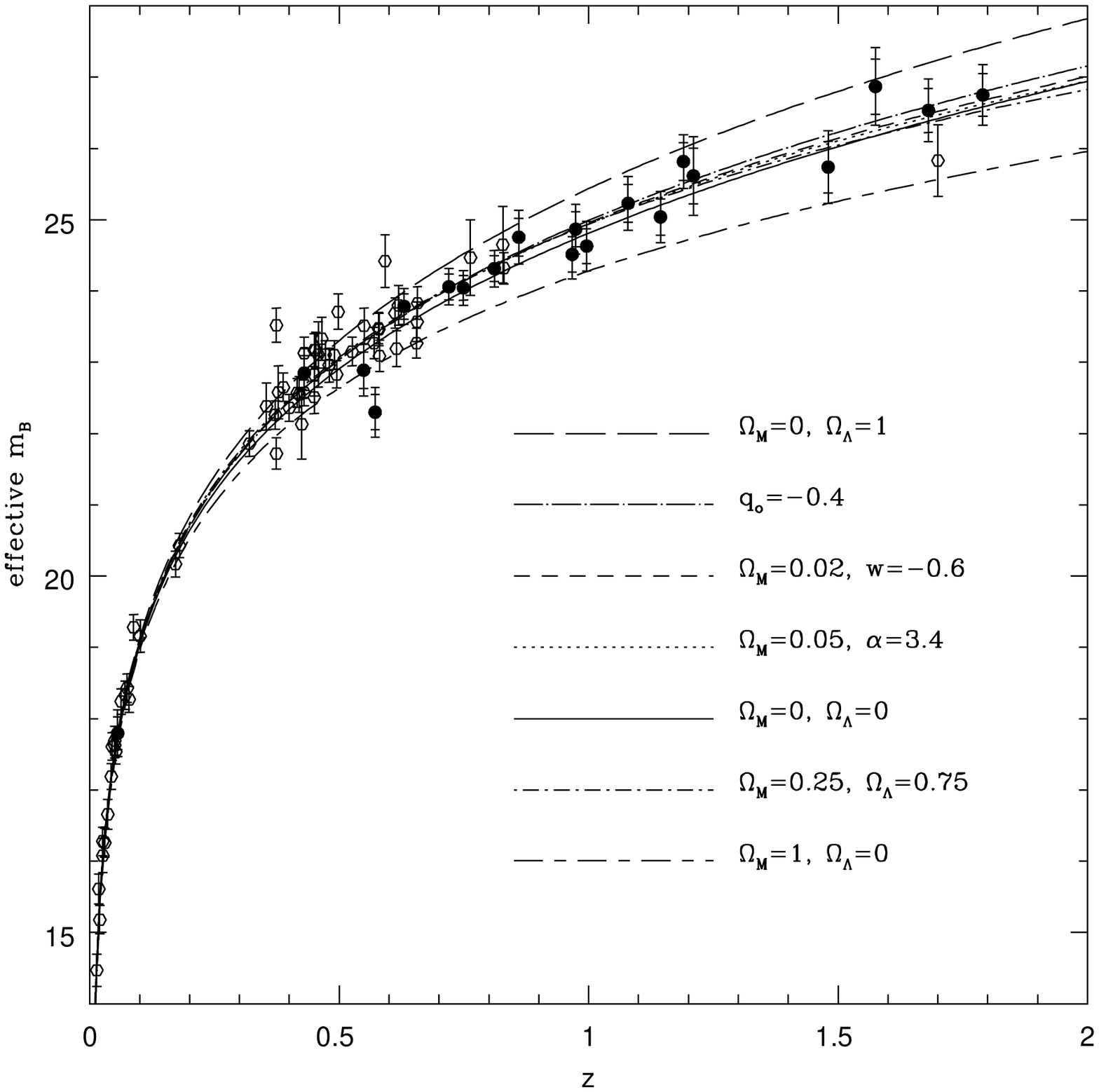,height= 15.0cm, width=15.0cm}
\caption{ Hubble diagram, taken from the supernova cosmology project. 
{\it ``The curve for
$(\Omega_\Lambda, \Omega_M) = (0,0)$ is practically identical to the best fit 
unconstrained cosmology''} \cite{perl}.}
\label{fig1}
\end{figure}

\end{document}